\documentclass[traditabstract,epsf]{aa}
\usepackage{epsfig}
\usepackage{txfonts}

\def\m2s2{\hbox{\,m$^{2}$\,s$^{-2}$}} 

\begin{document}

\title{Improved precision on the radius of the nearby super-Earth 55\,Cnc\,e\thanks{The photometric time series used in this work are available in electronic form at the CDS via anonymous ftp to cdsarc.u-strasbg.fr (130.79.128.5) or via http://cdsweb.u-strasbg.fr/cgi-bin/qcat?J/A+A/}}
\author{ Micha\"el Gillon$^{1}$, Brice-Olivier Demory$^{2}$, Bj\"orn Benneke$^2$, Diana Valencia$^2$, Drake Deming$^3$, Sara Seager$^2$, Christophe Lovis$^4$, Michel Mayor$^4$, Francesco Pepe$^4$, Didier Queloz$^4$, Damien S\'egransan$^4$, St\'ephane Udry$^4$}

\offprints{michael.gillon@ulg.ac.be}
\institute{
$^1$ Institut d'Astrophysique et de G\'eophysique,  Universit\'e de Li\`ege,  All\'ee du 6 Ao\^ut 17,  Bat.  B5C, 4000 Li\`ege, Belgium \\
$^2$ Department of Earth, Atmospheric and Planetary Sciences, Department of Physics, Massachusetts Institute of Technology, 77 Massachusetts Ave., Cambridge, MA 02139, USA\\
$^3$ Department of Astronomy, University of Maryland, College Park, MD 20742-2421, USA\\
$^4$ Observatoire de Gen\`eve, Universit\'e de Gen\`eve, 51 Chemin des Maillettes, 1290 Sauverny, Switzerland\\
}

\date{Received date / accepted date}
\authorrunning{M. Gillon et al.}
\titlerunning{Improved precision on the radius of the nearby super-Earth 55\,Cnc\,e}
\abstract{We report on new transit photometry for the super-Earth 55\,Cnc\,e obtained with {\it
Warm Spitzer}/IRAC at 4.5 $\mu$m. An individual analysis of these new data leads to a 
planet radius of  $2.21_{-0.16}^{+0.15}$ $R_\oplus$, which agrees well with the values 
previously derived from the MOST and {\it Spitzer} transit discovery data. 
A global analysis of both {\it Spitzer} transit time-series improves the precision on the radius of the planet 
at 4.5 $\mu$m to $2.20 \pm 0.12$ $R_\oplus$. 
We also performed an independent analysis of the MOST data, paying particular attention
to the influence of the systematic effects of instrumental origin on the derived parameters and errors by
including them in a global model instead of  performing a preliminary detrending-filtering processing.
We deduce  an optical planet radius of  $2.04 \pm 0.15$ $R_\oplus$  from this reanalysis of MOST data, 
which is consistent with the previous MOST result and with our {\it Spitzer} infrared radius.
Assuming the achromaticity of the transit depth, we performed a global analysis combining {\it Spitzer} 
and MOST data that results in a planet radius  of  $2.17 \pm 0.10$ $R_\oplus$ ($13,820 \pm 620$ km). 
These results point to 55\,Cnc\,e having a gaseous envelope overlying a rocky nucleus, in agreement
with previous works. A plausible composition for the envelope is 
water which would be in super-critical form given the equilibrium temperature of the planet.
\keywords{binaries: eclipsing -- planetary systems -- stars: individual: 55 Cnc - techniques: photometric} }


\maketitle

\section{Introduction}

Transiting planets are of fundamental interest for the field of exoplanetary science.
Their advantageous geometrical configuration relative to Earth enables the thorough
study of their physical, orbital and atmospheric properties, provided they orbit
around stars bright enough to permit high signal-to-noise follow-up observations.
For a given stellar type, this last condition is drastically more stringent for terrestrial planets
than for gas giants, leading to the conclusion that only the handful of solid planets 
that should transit stars of the closest solar neighborhood would be suitable for a thorough 
characterization with existing or future instruments (e.g. Seager et al. 2009).  

In this context, two teams, including ours,  independently announced the first transit detection 
for a solid planet orbiting a nearby star visible to the naked eye (Winn et al. 2011, hereafter W11;
Demory et al. 2011, hereafter D11).  This transiting `super-Earth', 55\,Cnc\,e, is 
the inner most of the five planets currently known to orbit around 55 Cancri, a G8-K0 dwarf 
star located at only 41 light-years from Earth (see D11 and references therein).  We detected  
one of its transits with  {\it Spitzer}, allowing us to deduce  a radius of $2.08_{-0.16}^{+0.17}$ 
$R_\oplus$ and  a mass of $7.81_{-0.53}^{+0.58}$ $M_\oplus$ for the planet (D11).  Together, these 
values favor a solid planet with a significant fraction of ice. On their side, Winn et al. detected 
several transits of 55\,Cnc\,e with the  MOST satellite and reported a planet radius consistent
 with ours, $2.00 \pm 0.14$ $R_\oplus$ (W11). Soon after this double transit detection, a third team reported 
impressively precise values for the host star's parameters based on new interferometric observations 
(von Braun et al. 2011, see Table 1). Thanks to these last results, made possible by the brightness
 of the star, our knowledge of the mass and size of 55\,Cnc\,e is only limited by the precision of 
 the radial velocities and transit photometry gathered so far.
 
Aiming to pursue the characterization of this fascinating planet, we monitored 
another of its  transits with {\it Spitzer} in our program dedicated to the search of nearby transiting 
low-mass planets (ID 60027). In the next two sections, we present these new data and their 
analysis, including a global analysis of MOST and {\it Spitzer} photometric time-series aiming to determine the size of the planet as precisely as possible. We discuss our results and their implications
in the last section of the paper. 

\section{New {\it Warm Spitzer} transit photometry}

We monitored 55\,Cnc with {\it Spitzer} on 20 June 2011 from 09h08 to 15h02 UT, corresponding to
a transit window of 55\,Cnc\,e as computed from our transit ephemeris presented in D11. The 
IRAC detector acquired 6230 sets of 64 subarray images at 
4.5 $\mu$m with an integration time of 0.01s. These 6230 sets were calibrated by the {\it Spitzer} 
pipeline version S18.18.0 and are available on the  {\it Spitzer} Heritage Archive 
database\footnote{http://sha.ipac.caltech.edu/applications/Spitzer/SHA} 
under the form of Basic Calibrated Data (BCD) files. Our reduction of these data was identical 
to the one presented in D11 and we refer the reader to this paper for details. Fig.~1 shows the 
resulting raw light curve composed of 6197 flux measurements, and also the time-series for the
 background counts and the $x$ and $y$ positions of the target's point-spread function (PSF) on 
 the detector array. We can notice from Fig.~1 that the background counts remained stable during 
 the whole run, unlike during our first transit observation (see D11, Fig.~1). 

\begin{figure}
\label{fig:1}
\centering                     
\includegraphics[width=9cm]{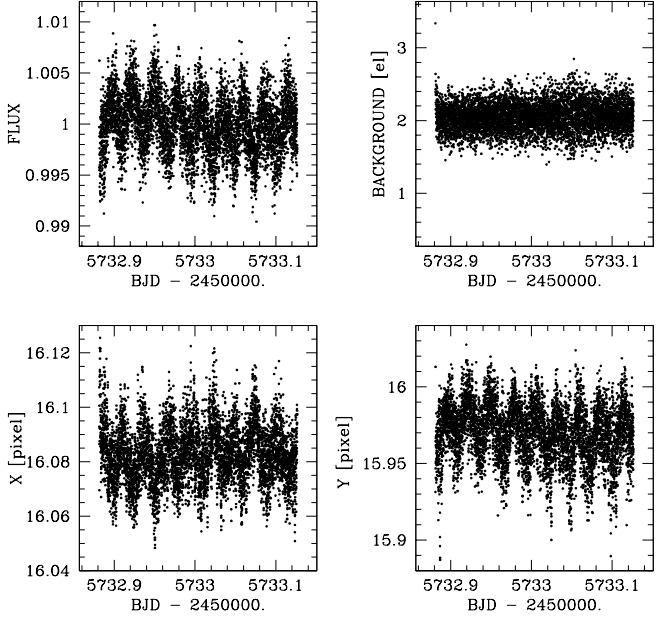}
\caption{{\it Top left}: raw light curve obtained for 55\,Cnc with {\it Warm Spitzer} at 4.5 $\mu$m. 
{\it Top right}: corresponding background time-series. {\it Bottom}: corresponding 
time-series for the $x$ ({\it left}) and 
$y$ ({\it right}) positions of the stellar center. The correlation between measured stellar counts and the 
stellar image  positions  is clearly noticeable. This `pixel-phase' effect is well-known for the 
InSb {\it Spitzer}/IRAC detectors (e.g. Knutson et al. 2008).
}
\end{figure}

\section{Data analysis}

\subsection{Individual analysis of the new {\it Warm Spitzer} data}

In a first step, we performed an individual analysis of our new  data. We used for 
this purpose our adaptative Markov-Chain Monte Carlo (MCMC) code 
(see D11 and references therein for details). Gaussian priors assumed for the
stellar parameters are shown in Table 1. Uniform priors were assumed for the other
parameters of the system.  First, we performed a thorough model comparison, performing for each 
tested model a MCMC analysis composed of one chain of 10,000 steps and deriving
the model marginal likelihood from the MCMC outputs using the method described
by Chib \& Jeliazkov (2001).  Each  model was composed of  a baseline representing the 
low-frequency instrumental and stellar effects multiplied by a  transit model 
computed under the formalism of Mandel \& Agol (2002) assuming a quadratic limb-darkening 
law. The baseline models included an  $x$- and $y$-position polynomial 
(D11, eq. 1) to model {\it Warm Spitzer} `pixel-phase' effect (see Fig.~1), which we added as needed
 to one or several functions of time. At the end, more than 30 models were tested, the one corresponding 
to the highest marginal likelihood has as baseline  a second-order position polynomial 
added to a fourth-order time polynomial and to a sinusoid. Because the Bayes factor (e.g. Carlin \& Louis 2008) between this model and the second most likely model is about 100, we selected it for sampling the posterior 
 probability density distributions of the transit parameters. We performed  a new MCMC 
 analysis for this purpose composed of two chains of 100,000 steps, and checked their convergence 
 using the statistical test of Gelman \& Rubin (1999). Table 2 provides the resulting values and errors for the transit parameters, while Fig.~2 shows the best-fit global model superimposed on the data and the 
best-fit transit model superimposed on the data divided by the best-fit baseline model. 

Comparing the resulting transit parameters shown in Table 2 to those derived in D11, 
we notice that the derived transit depths agree at better than $1 \sigma$ ($463_{-57}^{+64}$ ppm
 here $vs$ $410 \pm 63$ ppm for D11) but that the agreement for the derived impact parameters
  is only of $\sim$2.8 $\sigma$ ($0.509_{-0.074}^{+0.056}$ here $vs$ $0.16_{-0.13}^{+0.10}$  for D11).
 We suspect that this discrepancy comes mostly from the instrumental effect that affected
 our first transit data. Indeed, in D11 we had to model a sharp increase of the 
 effective gain of the detector during the run that was correlated to a strange behavior of 
 the background counts. We encountered this systematic effect  in 
 other {\it Warm Spitzer} time-series acquired in our program 60027 (Gillon et al., in prep.). 
Unfortunately, this effect occurred just during the egress of the transit, and we suspect that
it could have biased the derived marginalized posterior distribution function for the
transit impact parameter. 

The derived period and amplitude for the sinusoid function of the baseline model are
$59 \pm 2$ min and $107 \pm$ 24 ppm, consistent with the values derived in D11 for the first
 {\it Spitzer} transit, $51 \pm 3$ min and $115 \pm 27$ ppm. The origin of this low-amplitude
 periodic effect is still unclear (see D11). 

\begin{figure}
\label{fig:2}
\centering                     
\includegraphics[width=9cm]{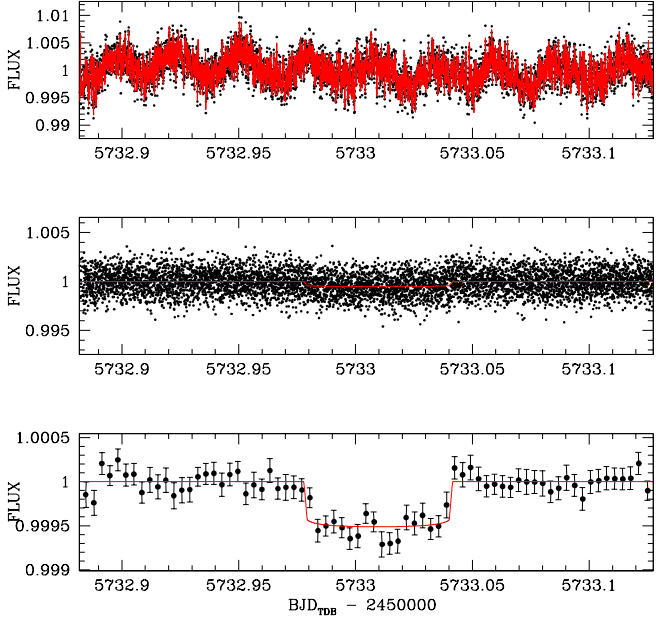}
\caption{$Top$: raw light curve for the second transit of 55\,Cnc\,e observed 
by {\it Warm Spitzer}, with the best-fit global model superimposed. $Middle$:
same light curve after division by the best-fit baseline model, with
the best-fit transit model superimposed. $Bottom$:
idem after binning per intervals of 5 min.}
\end{figure}

\subsection{Global analysis of {\it Warm Spitzer} data}

Aiming to minimize the impact of {\it Warm Spitzer} instrumental effects and to obtain 
the strongest constraints on the planet's infrared radius, we then performed a global MCMC analysis 
using as input data the two {\it Spitzer} light curves. We assumed for both  of them
the same baseline model as used for their individual analysis, with the exception that we did not 
fix the start time of the effective gain increase for  the first transit, but instead let it  be a free 
parameter to take into account our limited understanding of this systematic effect. 
The  Gaussian prior $P = 0.7365437 \pm 0.000052$ days based on the RV data analysis 
presented in D11 was used for the orbital period of the planet. The MCMC was composed
of two chains of 100,000 steps.

The best-fit transit model and the detrended light curves are visible  in Fig.~3. Table 2 presents 
the resulting values and errors for the transit parameters. We notice from this table that including
 the first transit as input data  improves the global precision on the planet's size, but not on the 
 transit duration/impact parameter, because of the degeneracy between the transit duration and 
 the increase of the effective gain that affected the first time-series. 
 
 \subsection{Independent analysis of the MOST data}
  
To assess the consistency of our {\it Spitzer} data with the  MOST
photometry, we performed our own analysis of the  MOST time-series based on
a different strategy from that used by W11. These authors analyzed the MOST data in two steps. 
First, they corrected their photometry from several instrumental effects known to affect MOST 
photometry. In a second step, they folded the corrected light curve on the orbital period of 55\,Cnc\,e, binned it to time intervals
of 2 min, and performed a MCMC analysis of the resulting light curve. To be consistent
with our analysis of the {\it Spitzer} data, we decided on our side to use the raw MOST
light curve (J. Winn, private com.) as input data and to include a model for the systematic effects
affecting MOST photometry in our global modeling. This strategy has the advantage to ensure 
the proper error propagation from the baseline model parameters to the transit parameters while avoiding any bias 
of a preliminary detrending/pre-whitening process on the derived results.

We thus started from the 27,950 photometric measurements gathered by MOST 
between 07 Feb 2011 00h15 to 22 Feb 2011 00h05 UT. For each photometric measurement, we also
had the corresponding background counts, PSF center position on the MOST CCD, a calculated
magnetic field, and several other external parameters. Following W11, we rejected measurements
obtained with a different exposure time from the bulk of the data, 41.82 s. We also rejected measurements
with an  $x$- or $y$-position off by more than 0.5 pixel from the median of the corresponding 
distributions. We indeed noticed a strong dependance of the measured fluxes to the PSF center positions, so we
chose to discard  discrepant measurements in terms of position to improve our chances to 
model the position effect with a simple analytical function. We also rejected outlier measurements
with a magnitude off by more than 0.01 mag compared to the median value. We then tested several 
models to represent  the data satisfactorily, our goal being at this stage not to find an optimum model but to 
identify discrepant measurements caused by transient effects (e.g. cosmic rays).  Once we had selected such a model and fitted it to the data, we analyzed the residuals and performed a $10$-sigma clipping to reject outliers. 
The resulting light curve was then binned with intervals of 5 min for the sake of computational speed, resulting
in 3,232 flux measurements.

We then began to iterate on models to represent  the data in the best way.
We noticed that the correlation of the fluxes and the positions could be satisfactorily modeled 
by an $x$- and $y$-position polynomial. 
We also noticed a strong dependence of the measured fluxes with the background counts and the magnetic field which we were also able to  model with polynomial functions of these external parameters.
Once detrended from the position, magnetic field and background effects, the data revealed 
after a Lomb-Scargle periodogram analysis (Press et al. 1992) significant power peaks at $\sim$1.7 hour, 
$\sim$0.94 days and $\sim$0.72 days, with false-alarm probabilities $< 10^{-11}$ for the three periods.
The first period corresponds to the orbital period of the MOST satellite 
(W11), while the second and third are close  to the rotation period of the Earth and, interestingly,
 to the orbital period of 55\,Cnc\,e, respectively. After  filtering the residuals from these periodic signals 
 by fitting sinusoids at the corresponding periods, the resulting light curve showed power excess at lower frequencies. We averaged them by dividing the light curve  into five light curves covering nearly equal durations, which we analyzed globally, assuming a different baseline for each of them .

We then performed a thorough model comparison as described above for {\it Spitzer} data to select 
the best model to represent the five resulting light curves. The baseline model that we selected 
contained a  third-order $x$- and $y$-position polynomial,
a  fourth-order background polynomial, a  fourth-order magnetic field polynomial, a fourth-order 
time polynomial, one sinusoid at the satellite orbital period and one at a period of $\sim$0.94 days. 
The transits of 55\,Cnc\,e were modeled with the  formalism of Mandel \& Agol (2002) assuming a 
quadratic limb-darkening law and using the Gaussian priors on the coefficients $u_1$ and $u_2$ shown in 
Table 1. In addition, we also introduced a model for the flux modulation 
at the orbital period of 55\,Cnc\,e. We represented this `phase curve' with a simple model dividing 
the planet into four equal slices and assuming for each of them a uniform luminosity and a 1:1 spin-orbit resonance 
for the planet. For each slice, the flux modulation was modeled by a simple sinusoid with its 
maximum corresponding to the center of the slice pointing toward Earth. We emphasize
 that our goal here was to select a model  that satisfactorily represents the flux modulation without 
 worrying about its physical relevance. Indeed, the amplitude of the observed flux modulation is much 
 too large to be attributed to the variable illuminated fraction of the planet or to its thermal emission, 
 as outlined by W11.  We tested introducing an occultation in our global model, but it did not improve the 
model marginal likelihood, consequently we discarded it from our final modeling. 

In our  complex model, the best-fit residuals of the five light curves still show a small
amount of correlated noise that we took into account in the same way as for {\it Spitzer} data 
by rescaling the measurements errors. At this stage, we performed a global analysis of our five light curves
and performed for this purpose two MCMC chains of 100,000 steps, whose convergence we successfully checked for the 
transit parameters using the statistical test of Gelman \& Rubin (1999). Table 2 provides the resulting
values and 1-$\sigma$ error bars for the transit and physical parameters, while Fig.~4 displays the best-fit models 
(global, phase curve + transits, transit) superimposed on the corresponding light curves. 

We notice that our results are consistent  with W11's results, the agreement being 
at better than 1-$\sigma$ for the transit depths and $\sim$ 1.5-$\sigma$ for the transit impact
parameters. Considering the large difference between both analyses, notably the different treatment of
 systematic effects, this agreement is reassuring for the robustness of the resulting inferences. 
 As described above, we divided the MOST data into five light curves, which were treated separately 
 from the others in our global MCMC analysis, except for the transit parameters. The free parameters
 for the phase curve model, i.e. the amplitudes of the sinusoid of each planet slice, were thus different for
  the five light curves. Interestingly, Fig. 4 shows that the five best-fit phase curve models 
  show some variability (see discussion  in Sec. 4.2). 
 
\subsection{Global analysis MOST + {\it Spitzer} data}

As can be noticed in Table 2, our independent analysis of {\it Spitzer} and MOST data led
to consistent results and similar precisions on the planet radius. For both instruments, 
the limiting factor on the transit depth precision is not the white noise associated with the flux
measurements but the high level of systematic effects that affect the photometry.
Aiming to minimize the impact of these systematics and to improve the precision 
on the transit's shape and the planet's size even more, we performed a global analysis of 
MOST and {\it Spitzer} data, assuming that the transit depth is exactly the same in both channels, 
i.e. that chromatic atmospheric transmission effects are not significant at this level of photometric precision.
This is an entirely reasonable assumption, considering the expected small atmospheric scale height 
of the planet (see Sec. 4.2 for more details).
The used priors, baseline models and analysis details were the same as in the separate 
analysis described above. Table 2 presents the resulting values and 1-$\sigma$ error bars
for the transit and physical parameters.

Finally, we assessed the validity of the assumption that both channels probe the 
same transit depth by performing a new global analysis of MOST and {\it Spitzer} data, 
this time adding as free parameter a difference in transit depth between the two
instruments. The derived transit depth difference {\it Spitzer} - MOST was 94 $\pm$ 80 ppm, i.e. 
the infrared radius of the planet is consistent with the optical radius, as already deduced
from the individual analysis of the data of both instruments. The other deduced
transit parameters were very similar to those shown in Table 2.

\begin{figure*}
\label{fig:1}
\centering                     
\includegraphics[width=18cm]{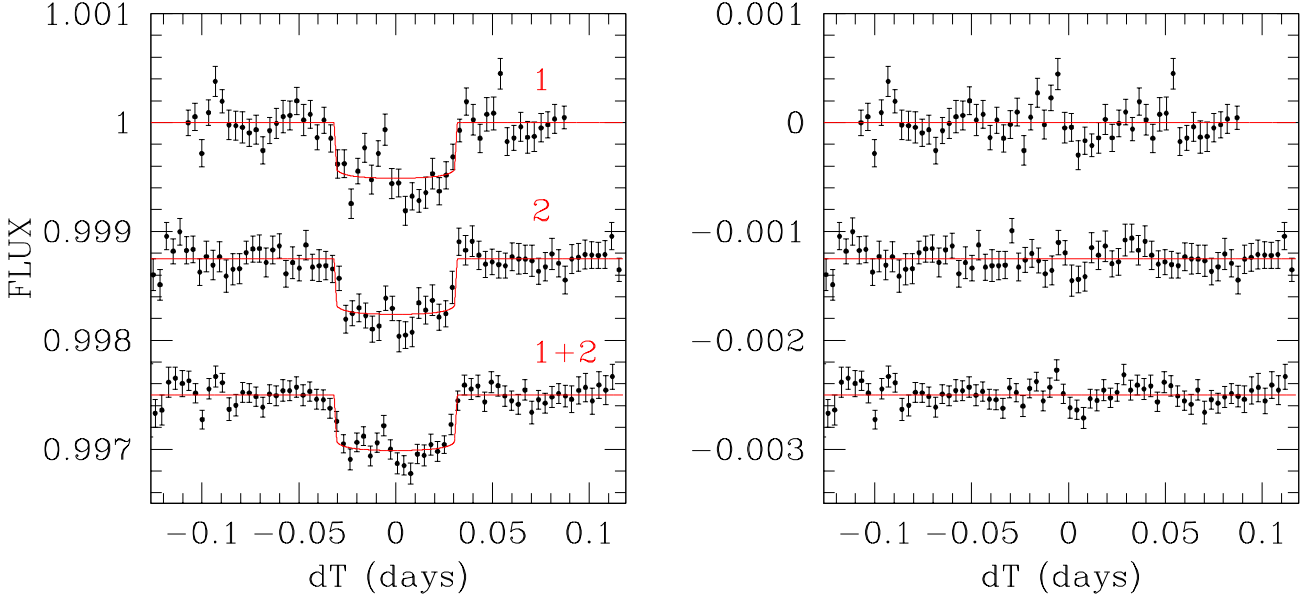}
\caption{$Left$: {\it Warm Spitzer} 55\,Cnc light curves (1 and 2: individual transits, 1+2:
combined light curve) divided by their best-fit baseline models
deduced from their global MCMC analysis, binned
to intervals of  5 minutes, with the best-fit transit model superimposed. $Right$:
residuals of the fit binned to intervals of  5 minutes. For both panels, two time-series
were shifted along the $y$-axis for the sake of clarity.}
\end{figure*}

\begin{table}
\begin{center}
\begin{tabular}{cc}
\hline \noalign {\smallskip}
Mass [$M_\odot$]                                                                & $0.905 \pm 0.015^1$                 \\ \noalign {\smallskip} 
Radius [$R_\odot$]                                                              & $0.943 \pm 0.010^1$                 \\ \noalign {\smallskip}  
$T_{eff}$ [K]                                                                         & $5196 \pm 24^1$                       \\ \noalign {\smallskip} 
Metallicity [Fe/H]  [dex]                                                         & $+0.31 \pm 0.04^2$                   \\ \noalign {\smallskip} 
Limb-darkening linear parameter $u_{1, 4.5 \mu m}$                              & $0.0711 \pm 0.0009^3$              \\ \noalign {\smallskip} 
Limb-darkening quadratic parameter $u_{2, 4.5 \mu m}$                        & $0.1478 \pm 0.0020^3$              \\ \noalign {\smallskip} 
Limb-darkening linear parameter $u_{1, MOST}$                                   & $0.657 \pm 0.090^4$                   \\ \noalign {\smallskip} 
Limb-darkening quadratic parameter $u_{2, MOST}$                            & $0.115 \pm 0.045^4$                    \\ \noalign {\smallskip} 
\hline \noalign {\smallskip}
\end{tabular}
\caption{Gaussian priors assumed in this work for the stellar parameters. $^1$von Braun et al. 2011, $^2$Valenti \& Fischer 2005, $^3$Claret \& Bloemen 2011, $^4$ W11.}
\end{center}
\end{table}

\begin{table*}
\begin{center}
{\scriptsize
\begin{tabular}{cccccc}
\hline \noalign {\smallskip}
Parameter & {\it Spitzer} transit 2 & {\it Spitzer} transit 1 \& 2   & MOST & {\it Spitzer} + MOST & Unit \\ \noalign {\smallskip} 
\hline \noalign {\smallskip}
Transit timing $T_{tr}$ &  $5733.0094_{-0.0011}^{+0.0012}$ & $5733.0085_{-0.0014}^{+0.0011}$   & $5607.0584_{-0.0017}^{+0.0016}$ &  $5733.0087_{-0.0011}^{+0.0013}$ & BJD$_{TDB} - 2450000$ \\ \noalign {\smallskip}
Orbital period  $ P$   &  0.7365437 (fixed)      & $0.7365460_{-0.0000046}^{+0.0000049}$     &  $0.7365437 \pm 0.0000052$ & $0.7365449_{-0.0000050}^{+0.0000046}$ & days \\ \noalign {\smallskip}                            
Transit depth $(R_p/R_\ast)^2$                & $463_{-54}^{+57}$& $458 \pm 47$  & $394_{-51}^{+61}$ & $447_{-38}^{+40}$ & ppm         \\ \noalign {\smallskip} 
Planet-to-star radius ratio $(R_p/R_\ast)$             & $0.0215 \pm 0.0013$  &$0.0214 \pm 0.0011$  & $0.0198_{-0.0013}^{+0.0015}$ & $0.02113_{-0.00091}^{+0.00093}$ &   \\ \noalign {\smallskip} 
Transit circular impact parameter $b$  &  $0.509_{-0.074}^{+0.056}$     & $0.500_{-0.085}^{+0.057}$   & $0.44_{-0.16}^{+0.11}$ & $0.459_{-0.084}^{+0.076}$ & $ R_*$        \\ \noalign {\smallskip} 
Transit duration $W$     &  $0.0589_{-0.0023}^{+0.0026}$  & $0.0593_{-0.0023}^{+0.0029}$   & $0.0612 \pm 0.0039$ &  $0.0607_{-0.028}^{+0.025}$&  days      \\ \noalign {\smallskip} 
Orbital inclination $i$    & $81.7_{-1.0}^{+1.2}$      & $81.8_{-1.0}^{+1.4}$  & $82.8_{-1.8}^{+2.6}$  & $82.5_{-1.3}^{+1.4}$ & deg                \\ \noalign {\smallskip} 
Planet radius  $ R_p$                                      &  $2.21_{-0.16}^{+0.15}$ & $2.20 \pm 0.12$   &  $2.04_{-0.14}^{+0.15}$ & $2.173_{-0.098}^{+0.097}$ & $R_{\oplus}$   \\ \noalign {\smallskip} 
\hline \noalign {\smallskip}
\end{tabular}}
\caption{Median and 1-$\sigma$ limits of the marginalized posterior distributions obtained for the parameters of 55\,Cnc\,e from our MCMC analysis of the {\it Warm Spitzer} data.}
\end{center}
\end{table*}

\begin{figure*}
\label{fig:4}
\centering                     
\includegraphics[width=18cm]{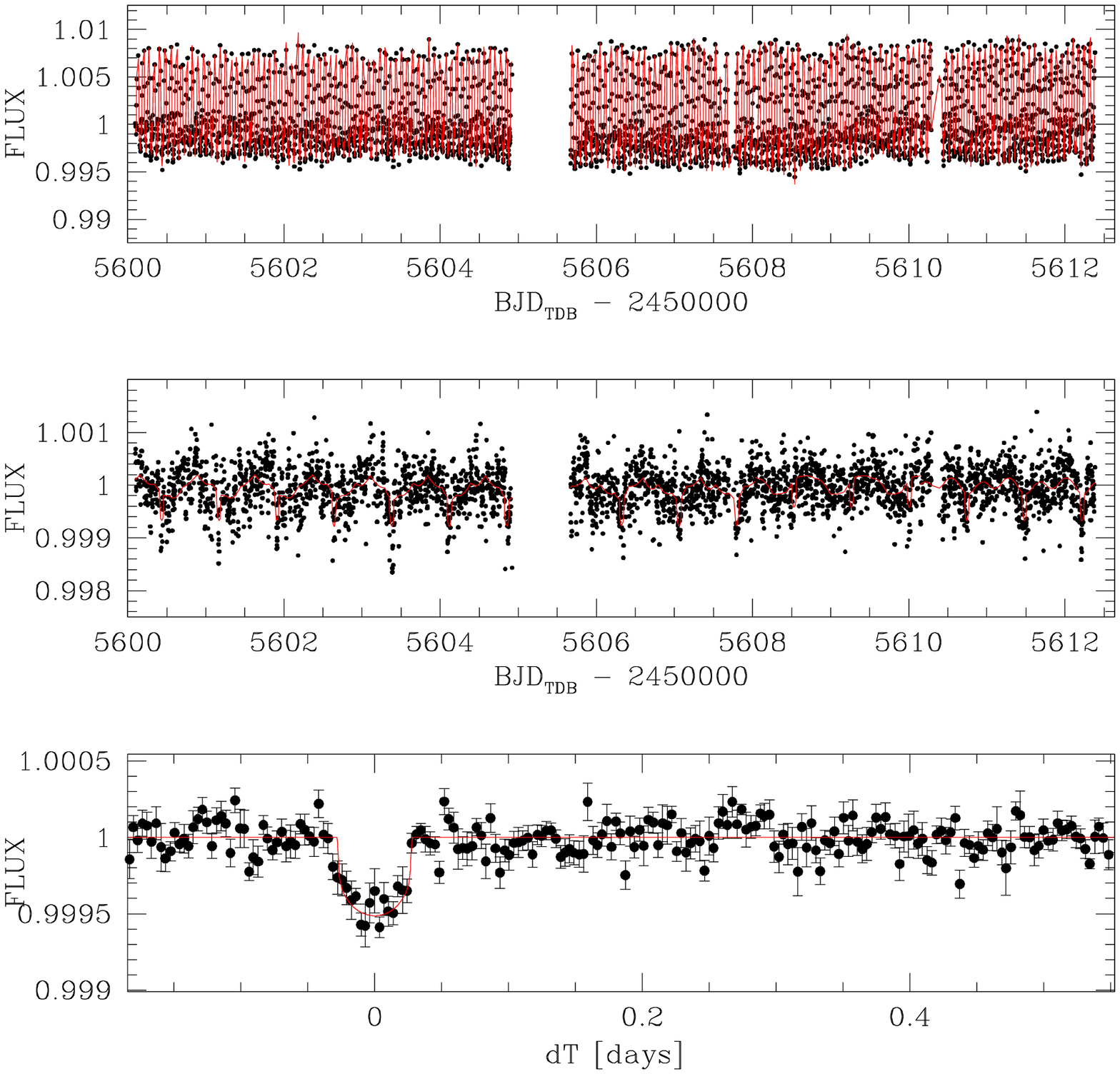}
\caption{$Top$: 55\,Cnc  MOST raw photometry with our best-fit global model superimposed. $Middle$: MOST photometry divided by the best-fit baseline model, and with the best-fit transit + phase-curve model superimposed. $Bottom$: MOST photometry divided by the best-fit baseline + phase curve model, folded with the best-fit orbital period of 55\,Cnc\,e and binned per 5 min intervals, with the best-fit transit model superimposed.}
\end{figure*}

\section{Discussion}

\subsection{The composition of 55\,Cnc\,e}

To infer the composition of the planet we used the internal structure
model by Valencia et al. (2006, 2010) and considered representative
compositions for rocky and volatile planets. The range in radii for
rocky planets are delimited by a pure iron composition that yields
the smallest radius and a magnesium-silicate oxide composition (devoid
of iron) that yields the largerst radius. Owing to the fact that iron,
magnesium, and silicate are all refractory elements with similar condensation
temperatures, planets are unlikely to form with either of these extreme
compositions. Two plausible compositions we looked at are Earth-like
composition (33\% iron core above a silicate mantle with 10\% of iron
and 90\% of silicate by mol) and an iron-enriched composition (63\%
iron core above a Mg-Si mantle, with no iron). 

We find that 55\,Cnc\,e is too large to be made out of just rocks despite
its relative high bulk density of $\rho=4.0_{-0.3}^{+0.5}$ g cm$^{-3}$ (e.g.
Earth's bulk density is $\rho_{\oplus}=5.5$ g cm$^{-3}$) obtained with the
radius reported in this study. Therefore, it has to have an envelope
of volatiles. We consider two compositions for the gaseous envelope:
a hydrogen and helium mixture, and a pure water vapor composition
(which at these temperatures is super-critical). We added different
amounts of envelope to an Earth-like nucleus. The results show (see
Fig.~5) that the data may be fitted with a H-He envelope
of $\sim$0.1\% by mass or a H$_{2}$O envelope of $\sim$20\% by
mass. As described in D11, based on simple atmospheric
escape calculations described in Valencia et al. (2010), this low-mass
envelope of H-He would escape in Myr timescales, whereas a water-vapor
envelope would escape in billions of years timescales. Thus, the favored
composition for 55\,Cnc\,e is a volatile planet with a water dominated
envelope comprising tens of percent of the total mass of the planet. 

The radius obtained in this study is larger than that reported by
D11 and above a one-sigma level of W11
value, ruling out a rocky composition for 55\,Cnc\,e. An interesting
characteristic of this planet is that it has an intermediate composition
(see Fig.~6). While most of its mass is bound in  a rocky nucleus, it has a 
non-negligible (most likely high-molecular) envelope. This lies between the 
composition of GJ\,1214\,b, which probably consists
mostly of water (see Valencia et al. 2011), and the terrestrial
planets in our solar system and exoplanets such as `super-Mercuries' CoRoT-7\,b
(Hatzes et al. 2011) and Kepler-10\,b (Batalha et al. 2011). 

\begin{figure}[h]
\label{fig:5}
\centering                     
\includegraphics[width=9cm]{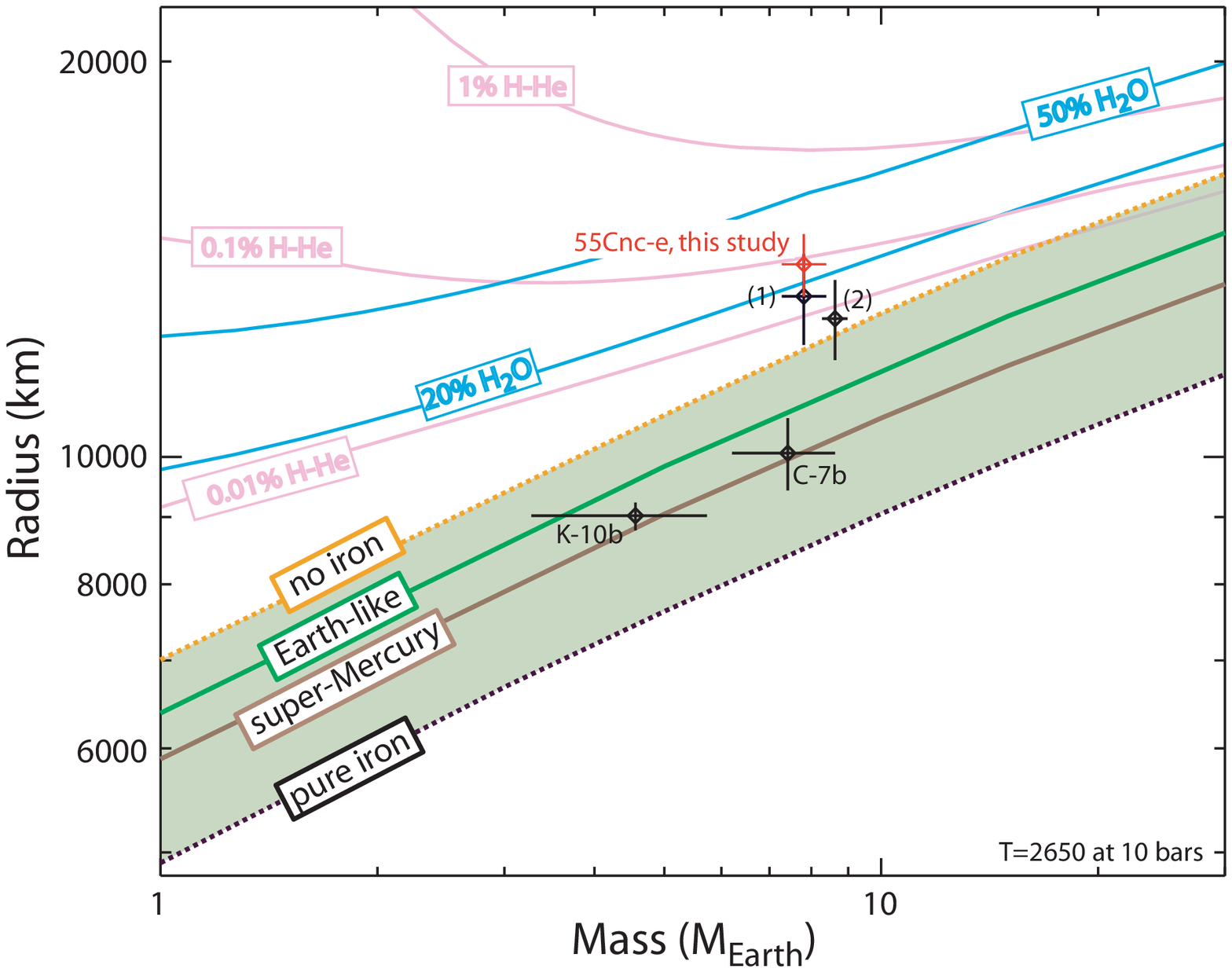}
\caption{Composition of 55\,Cnc\,e. Mass-radius relationships for four different
rocky compositions: pure iron, super-Mercury (63\% iron core above
a 37\% silicate mantle), Earth-like (33\% iron core above a 67\% silicate
mantle with 10\% iron by mol) and a silicate planet with no iron. 
The green band depicts the range of rocky compositions
of planets between 1 and 30 $M_{\oplus}$. Two families of volatile
planets are considered: envelopes of H-He with 0.01, 0.1 and 1\% by
mass (pink) and envelopes of water with 20, 50\% water. Kepler-10b
(K-10b) and  CoRoT-7b (C-7b) are shown for reference. Data for 55\,Cnc\,e
from this study are shown in red, and are taken from D11 (1) and W11 (2).}
\end{figure}

\begin{figure}[h]
\label{fig:6}
\centering                     
\includegraphics[width=8cm]{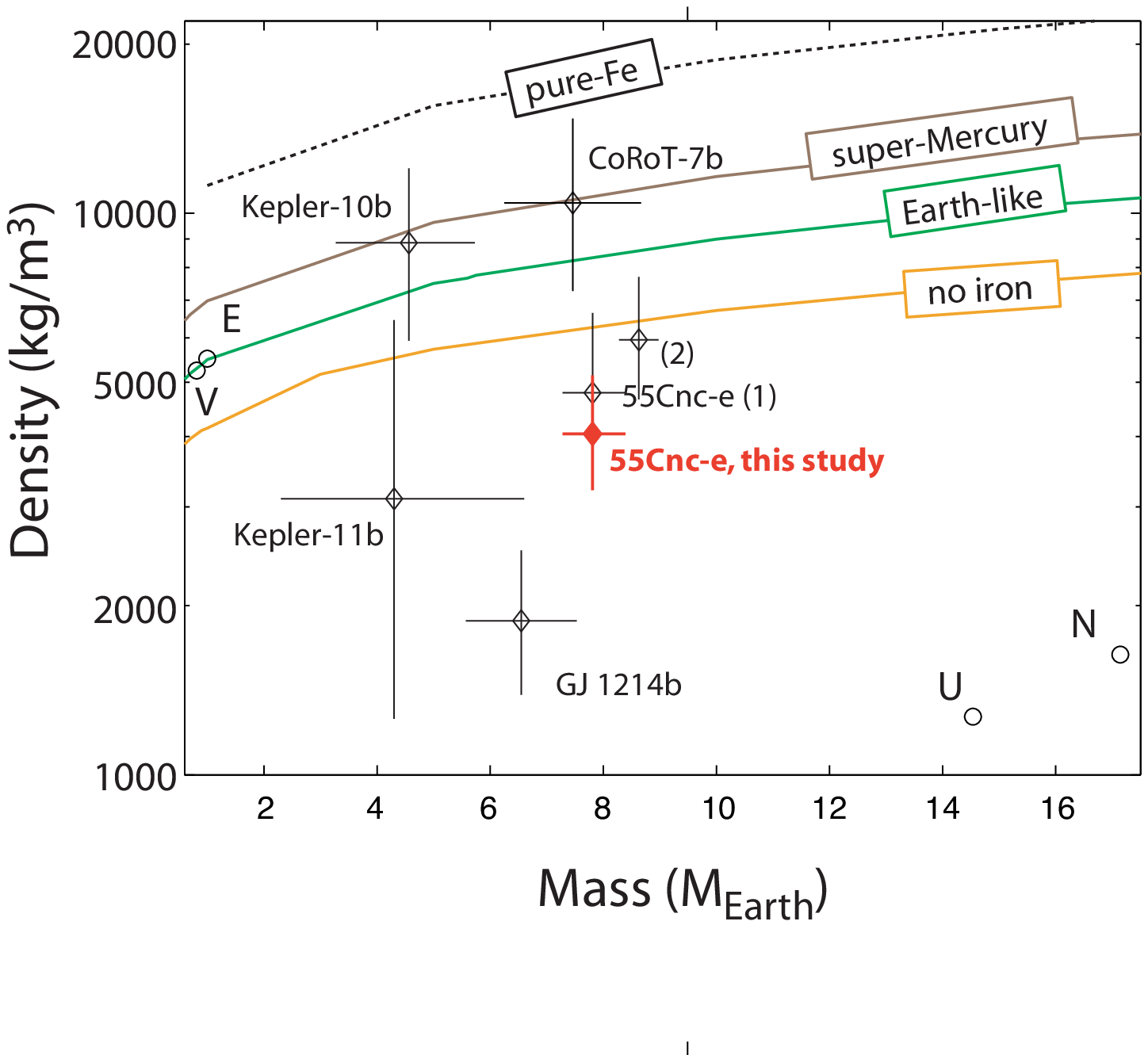}
\caption{Density-mass relationships of low-mass exoplanets. Four rocky compositions
are shown (see Fig.~5 for description). 55\,Cnc\,e's radius from
this study shows the planet cannot be rocky and instead is intermediate
between the envelope-rich planets (eg. GJ\,1214\,b) and the rocky ones.
Earth, Venus, Uranus, Neptune, Kepler-11\,b, GJ\,1214\,b, Kepler-10\,b, and
CoRoT-7\,b are shown for reference.}
\end{figure}

\subsection{Potential atmospheric studies of 55\,Cnc\,e}

Follow-up observations made to detect the spectral signature of the planetary atmosphere
may bring new constraints on the chemical composition of 55\,Cnc\,e. The planet is 
suitable for these observations, because it orbits around a very bright, nearby star at 
an extremely close distance. Still, depending on the nature of the planetary atmosphere,
follow-up observations can be extremely challenging due to the shallow transit depth.

Atmospheric modeling (Fig.~7) suggests that transmission signatures of an atmosphere
in hydrostatic equilibrium could be on the order of 100 ppm, if the
planet has been able to accrete and retain a hydrogen-dominated atmosphere.
Such signatures could be detected or excluded with currently available
space-based instrumentation.  However, the short evaporation timescale for
a H-He envelope strongly disfavors this scenario, as stated above. 
The most probable scenario implies an envelope mostly composed of water and other 
ices which would result in much weaker transmission signatures on the level of tens of ppm
due to the higher mean molecular mass (Fig.~7). A direct detection of such
a water/ices dominated atmosphere on 55\,Cnc\,e will probably have
to wait until next generation instruments on-board JWST become available.

Stronger transmission signatures are plausible, if one hypothesizes
that the planet is surrounded by a low-density halo of atomic gas
resulting from atmospheric escape. Both the small Roche lobe and the
high equilibrium temperature favor a strong atmospheric escape and
the lost atmospheric mass could  be readily replenished by evaporation
of the planet's surface or oceans (Yelle et al. 2008, Schaefer \& Fegley 2009,
Ehrenreich 2010). The evaporated gas should be readily
dissociated in the high-temperature halo. The resulting atomic species
like C, H, O, Na, Mg and Ca could be detectable in transmission
because of their strong absorption cross sections near their electronic
transition lines and the large extent of the halo (e.g. Mura et al. 2011). 
The exosphere of 55\,Cnc\,e could also explain the flux modulation
at the planet's orbital period detected by W11 and our own analysis of MOST data. 
Because the amplitude of this modulation is too large for the planet's thermal emission or 
reflected light, W11 hypothesized the induction by the planet of a patch of enhanced 
magnetic activity on the star. Another explanation is that a part of the gases 
escaped from the planet's atmosphere forms a circumstellar disk along the planet's orbit
similar to Io's donut-shaped cloud (e.g. Schneider \& Bagenal 2007). The short lifetime of these evaporated gases
would modulate the opacity of this cloud along the orbit. Furthermore, the produced ionic 
species would rotate with the star magnetic field, leading to a modulation of the 
resulting `phase curve' at the rotational period of the star, $42.7 \pm 2.5$ days (Fischer et al. 2008), 
which is consistent with the apparent variability of the fitted phase curve model in
our analysis (Fig. 4, middle panel). More work is needed to assess the plausibility
of this hypothesis.

Because of its extreme proximity to its host star and the resulting incident 
flux of $\sim$3.3 10$^9$ erg s$^{-1}$ cm$^{-2}$, 55\,Cnc\,e's thermal emission could
be measured through infrared occultation photometry.  
With realistic assumptions for the Bond albedo and heat distribution efficiency of
 the atmosphere, occultation depths ranging between 90 and 150 ppm are expected 
 at 4.5 $\mu$m. The photometric precision demonstrated here shows that a
  low-amplitude eclipse like this could be detected by {\it Spitzer}, provided several 
 events are monitored. This is the goal of our accepted {\it Spitzer} program 80231, 
 and we are waiting eagerly for these future observations to learn more
about this fascinating planet.

\begin{figure}
\label{fig:7}
\centering                     
\includegraphics[width=9cm]{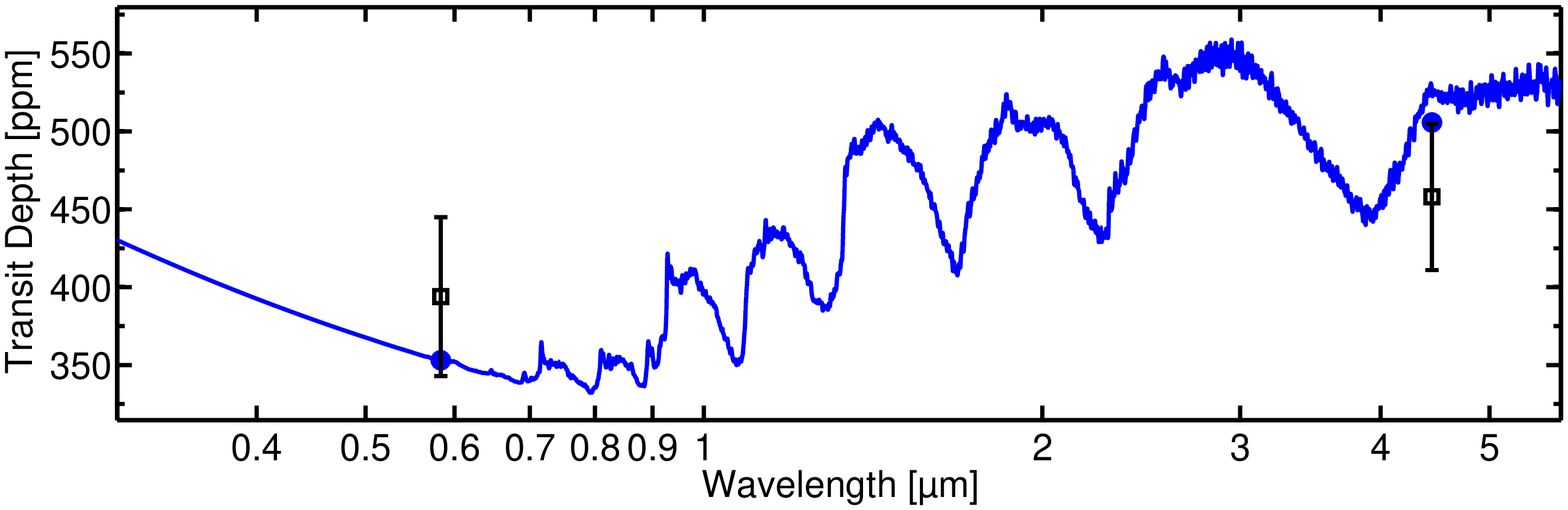}
\includegraphics[width=9cm]{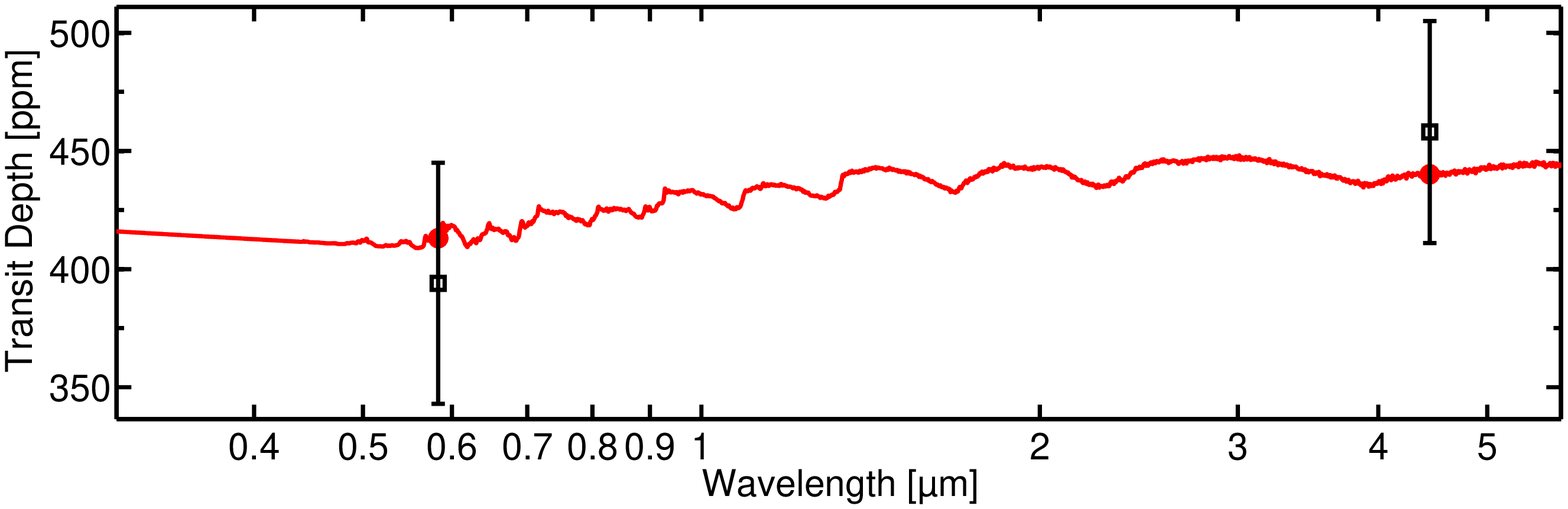}
\caption{Broadband transit depth measurements from MOST and {\it Spitzer} (squares) 
compared to models (lines + filled circles). Theoretical transmission spectra for 55\,Cnc\,e 
are displayed for an atmosphere composed of 100\% water vapor ($bottom$) and an 
atmosphere with solar composition ($top$).  The model assumes a gravitationally bound
 atmosphere in radiative-convective equilibrium (Benneke \& Seager, 2011, submitted).  }
\end{figure}

\begin{acknowledgements}
We are grateful to J. Winn for having provided us with the MOST photometry. 
This work is based in part on observations made with the {\it Spitzer Space Telescope}, which is operated 
by the Jet Propulsion Laboratory, California Institute of Technology under a contract with NASA. Support for 
this work was provided by NASA. We thank the {\it Spitzer} Science Center staff for efficient scheduling of 
our observations. M. Gillon is FNRS Research Associate. 
\end{acknowledgements} 

\bibliographystyle{aa}

\end{document}